\documentclass[a4paper,12pt]{article}
\usepackage[dvips]{graphicx}
\newlength{\dinwidth}
\newlength{\dinmargin}
\begin{document}
\begin{flushright}
  IC-HEP/99-03 \\
\end{flushright}

\vspace*{1cm}
\begin{center}
  {\huge \bf Future {\em ep} Physics:\\
       \em  The Outlook for HERA} \\
  \vspace*{1cm}
  {\large \sl K.R. Long\\
     Imperial College, London} \\
  \vspace*{1cm}
  {\sl On behalf of the H1 and ZEUS Collaborations}\footnote 
  {\sl Presented at the XIV$^{th}$ International Workshop on High Energy 
    Physics and Quantum Field Theory, Moscow, 27$^{th}$ May - 2$^{nd}$ June 1999.}
\end{center}

\begin{abstract}
\noindent
{\it 
The luminosity of the electron-proton collider, HERA, will be
increased by a factor of five during the long shutdown in the year 2000. 
At the same time longitudinal lepton beam polarisation
will be provided for the collider experiments H1 and ZEUS. These far
reaching upgrades to the machine will be matched by upgrades to the
detectors. The result will be a unique facility for the study of the
structure of the proton and the nature of the strong and electroweak
interactions. The physics potential of the upgraded accelerator is
discussed here together with a brief description of the HERA machine
and collider detector upgrades.
} 
\end{abstract}

\section{Introduction}
The electron-proton collider HERA started operation in the summer of
1992. The proton beam energy was 820 GeV while the electron beam
energy was 27.5 GeV. The $\sim$25 nb$^{-1}$ of data collected by the 
experiments ZEUS and H1 in the first running period extended the kinematic 
range covered by deep inelastic scattering, DIS, measurements by two orders
of magnitude in both $Q^{2}$, the four-momentum transfer squared, and $x$,
the fraction of the proton four-momentum carried by the struck
quark. The proton structure function was extracted from the data and
observed to rise rapidly with decreasing $x$ \cite{Abt,Derrick}; a 
dramatic result, not expected by many. 

In the years 1992-1997 H1 and ZEUS have each collected a luminosity of 
$\sim$1 pb$^{-1}$ using electron beams and $\sim$50 pb$^{-1}$ 
using positron beams. These data have been used to make measurements which 
test the electroweak Standard Model, SM, and the theory of the strong 
interactions, QCD, in both neutral current, NC, and charged current, CC, DIS. 
Jet analyses in DIS and photoproduction have been used to address fundamental
issues in QCD. The observation of diffraction in DIS has led to a
careful investigation of the transition from the kinematic region in
which perturbative QCD is valid to the region where phenomenological
models based on Regge theory must be applied (see for example \cite{Cooper} 
and \cite{Abramovicz} and references therein). 

During the running period August 1998 to April 1999 $\sim$20 pb$^{-1}$ 
of $e^{-}p$ data were delivered with a proton beam energy of 920 GeV. The data
collected by H1 and ZEUS will be used to study the dependence of the
NC and CC DIS cross sections on the charge of the lepton beam.

The HERA experiments will continue to take data until May 2000 when a
long, 9 month, shutdown is scheduled. The shutdown will be used to
upgrade the HERA accelerator and the collider detectors. The HERA
luminosity will be increased by a factor of five and longitudinal
lepton beam polarisation will be provided for ZEUS and H1. The physics
motivation for this major upgrade programme is discussed in detail in
reference \cite{Ingelman}. Reference \cite{Ingelman} also contains a 
discussion of the physics potential of HERA were polarised proton beams and 
beams of heavy nuclei to be provided. These interesting possibilities will 
not be discussed further here.

This report is organised as follows: section 2 contains a selection of
current results from H1 and ZEUS which will benefit from more data or
from improvements in the quality of the data expected to come from the
detector upgrades; the HERA machine and the collider detector upgrades
are reviewed in section 3; section 4 contains a selection of key
physics topics which will be addressed after the upgrade; finally,
section 5 contains a summary. 

\section{A Selection of Open Questions}

The wealth of data provided by the successful operation of HERA over
the past seven years has allowed a set of open questions to be
clearly defined. Since a discussion of each open question can not be
attempted here, the case for the upgrade of the accelerator and of the 
collider detectors will be made using three key examples. 

\subsection{Determination of the Gluon Density of the Proton}

The double differential cross section for $e^{\pm}p$ NC DIS may be written in
the form  
\begin{equation}
\frac{d\sigma^{\rm NC}_{e^{\pm}p}}{dxdQ^{2}} =\frac{2\pi\alpha^{2}}
{xQ^{4}} [Y_{+}F_{2}^{\rm NC} \mp Y_{-}xF_{3}^{\rm NC}
- Y^{2}F_{L}^{\rm NC}] = \frac{2\pi\alpha^{2}}{xQ^{4}} Y_{+}\tilde{\sigma}
^{\rm NC}_{e^{\pm}p}
\end{equation}
where $Y_{\pm} = 1 \pm (1-y)^{2}$ with $y=Q^{2}/ x s$, $\sqrt{s}$ is
the $e^{\pm}p$ centre of mass energy, $\alpha$ is the fine structure constant 
and $\tilde {\sigma}^{\rm NC}_{e^{\pm}p}$ is referred to as the reduced cross 
section. The structure functions $F_{2}^{\rm NC}$ and $xF_{3}^{\rm NC}$ 
contain parton density functions, PDFs, and electromagnetic and weak couplings
\cite{Ruckl}. The longitudinal structure function $F_{L}^{\rm NC}$ arises from QCD 
corrections to the naive quark parton model and makes a significant 
contribution to the NC cross section only at low $Q^{2}$. The structure 
function $F_{2}^{\rm NC}$ can be written as the sum of two terms 
$F_{2}^{\rm NC} = F_{2}^{\rm em} + F_{2}^{\gamma/Z}$. $F_{2}^{\rm em}$ 
contains the purely electromagnetic contribution, while $F_{2}^{\gamma/Z}$ 
receives contributions from the parity conserving terms which arise from $Z$ 
exchange and photon-$Z$ interference. ZEUS and H1 have measured the structure 
function $F_{2}^{\rm NC}$ over the kinematic range 
\( 3.6 \times 10^{-5} < x < 0.65 \) and \( 0.14 < Q^{2} < 2 \times
10^{4}\,{\rm GeV}^{2} \) and extracted $F_{2}^{\rm em}$ from these
measurements 
\cite{Aid,Breitweg}. The error in $F_{2}^{\rm em}$ is now dominated 
by systematic uncertainties over a large part of the this range. Fits
to the data using next to leading order, NLO, QCD have been used to
extract the gluon density, $xg$, with a precision varying from 20\% at
$x \approx 3 \times 10^{-5}$, $Q^{2}=20\,{\rm GeV}^{2}$ to 15\% at $x
\approx 4 \times 10^{-4}$, $Q^{2}=7\,{\rm GeV}^{2}$ \cite{ZEUS}. This
is an indirect determination of $xg$ since the positron scatters from
a quark. A quantity more directly sensitive to the gluon density is
the charm production cross section. This may be measured using a
sample of events which contain a $D^{\ast}$ meson reconstructed using
the decay chain $D^{\ast} \rightarrow D^{0}\pi \rightarrow
K\pi\pi$. Both H1 and ZEUS have performed such a measurement in 
DIS and in photoproduction \cite{67,Adloff}. A sensitive test of QCD 
can be made by comparing the measured charm cross sections to the cross 
sections obtained from the scaling violations of $F_{2}^{\rm em}$. At present 
the data are not sufficiently precise to allow a quantitative comparison to 
be made.

In order to investigate the extent to which the gluon distribution
extracted from the scaling violation of $F_{2}^{\rm em}$ agrees with that 
inferred from the charm production cross sections it is necessary to extend the
range of $x$ and $Q^{2}$ over which $F_{2}^{\rm em}$ is measured to high 
precision so increasing the precision with which $xg$ can be extracted. 
At the same time a much larger charm data set is required to reduce the 
errors on the charm cross sections. The HERA luminosity upgrade combined 
with the improved charm tagging efficiency to be provided by the detector
upgrades will make it possible to address the issue of whether the
gluon distribution extracted from $F_{2}^{\rm em}$ is also that required to 
describe the charm production cross sections.

\subsection{Neutral Current DIS Cross Section Measurement at High $Q^{2}$}

\begin{figure}
  \begin{center}
    \includegraphics[width=.45\textwidth]{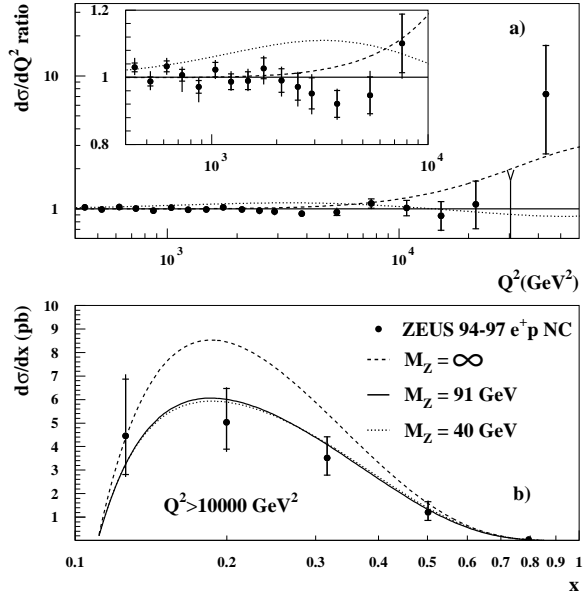}
  \end{center}
  \caption{
    (a) The points with error bars give the ratio of the 
    measured cross sections, $d \sigma^{\rm NC}_{e^{+}p}/dQ^{2}$,
    to the SM prediction using the CTEQ4D \protect{\cite{CTEQ4}} parton
    densities and fixing the Z mass, $M_{Z}$, at its nominal value of
    91.175~GeV. (b) $d \sigma^{\rm NC}_{e^{+}p}/dx$ for 
    $Q^{2} > 10000\,{\rm GeV}^{2}$. The three lines show the predictions
    of the SM for $M_{Z} = 91.175\,{\rm GeV}$, for $M_{Z} = 40\,{\rm
    GeV}$ (dotted line) and for no Z contribution (dashed line)
    \protect{\cite{DESY-99-056}}.
    \label{Fig:Fig1}
  }
\end{figure}

\begin{figure}
  \begin{center}
    \includegraphics[width=.45\textwidth]{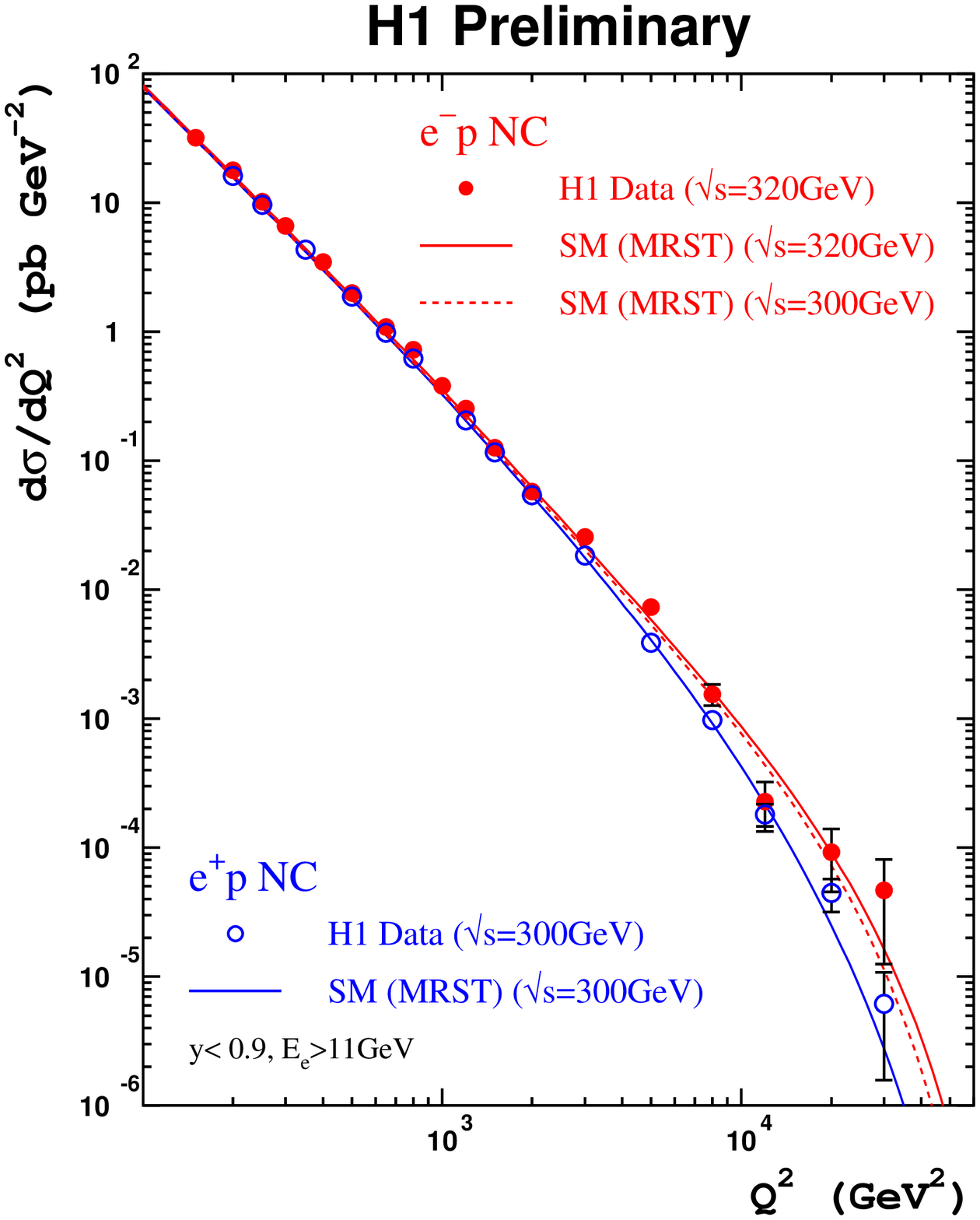}
  \end{center}
  \caption{
    Preliminary inclusive neutral current cross sections 
    as a function of $Q^{2}$ for H1 $e^{-}p$ data from 1998-99 (solid
    points) and for H1 $e^{+}p$ data from 1994-97 (open points)
    \protect{\cite{Chekelian}}.  The SM curves were evaluated using
    the MRST PDFs \protect{\cite{MRST}}.
    \label{Fig:Fig2}
  }
\end{figure}

The effect of $Z$ exchange on the NC DIS cross section has been observed
at high $Q^{2}$. This is demonstrated in figure \ref{Fig:Fig1} where the single
differential cross sections \( d \sigma^{\rm NC}_{e^{+}p}/dQ^{2} \) and 
\( d \sigma^{\rm NC}_{e^{+}p}/dx \) are compared with the expectations
of the SM \cite{DESY-99-056}. The SM, including the $Z$ exchange contribution, 
describes the data while a calculation in which the $Z$ exchange contribution
is ignored is in clear conflict with the data. The SM predicts that
the $\gamma-Z$ interference contribution to NC $e^{+}p$ scattering is 
destructive while in $e^{-}p$ scattering it is constructive. The measured 
single differential cross sections \( d \sigma^{\rm NC}_{e^{+}p}/dQ^{2} \) 
and \( d \sigma^{\rm NC}_{e^{-}p}/dQ^{2} \), shown in figure
\ref{Fig:Fig2}, are consistent 
with this expectation \cite{Chekelian}. It will be possible to extract the
structure function $xF_{3}^{\rm NC}$ by combining the $e^{-}p$ and $e^{+}p$ 
data, however, with the luminosities currently available the precision of the 
measurement will be dominated by the statistical uncertainty. To improve the
measurement will require a significant increase in luminosity and some
improvements in the detection of the scattered lepton at high $x$ and at
high $Q^{2}$.

\begin{figure}
  \begin{center}
    \includegraphics[width=.6875\textwidth]{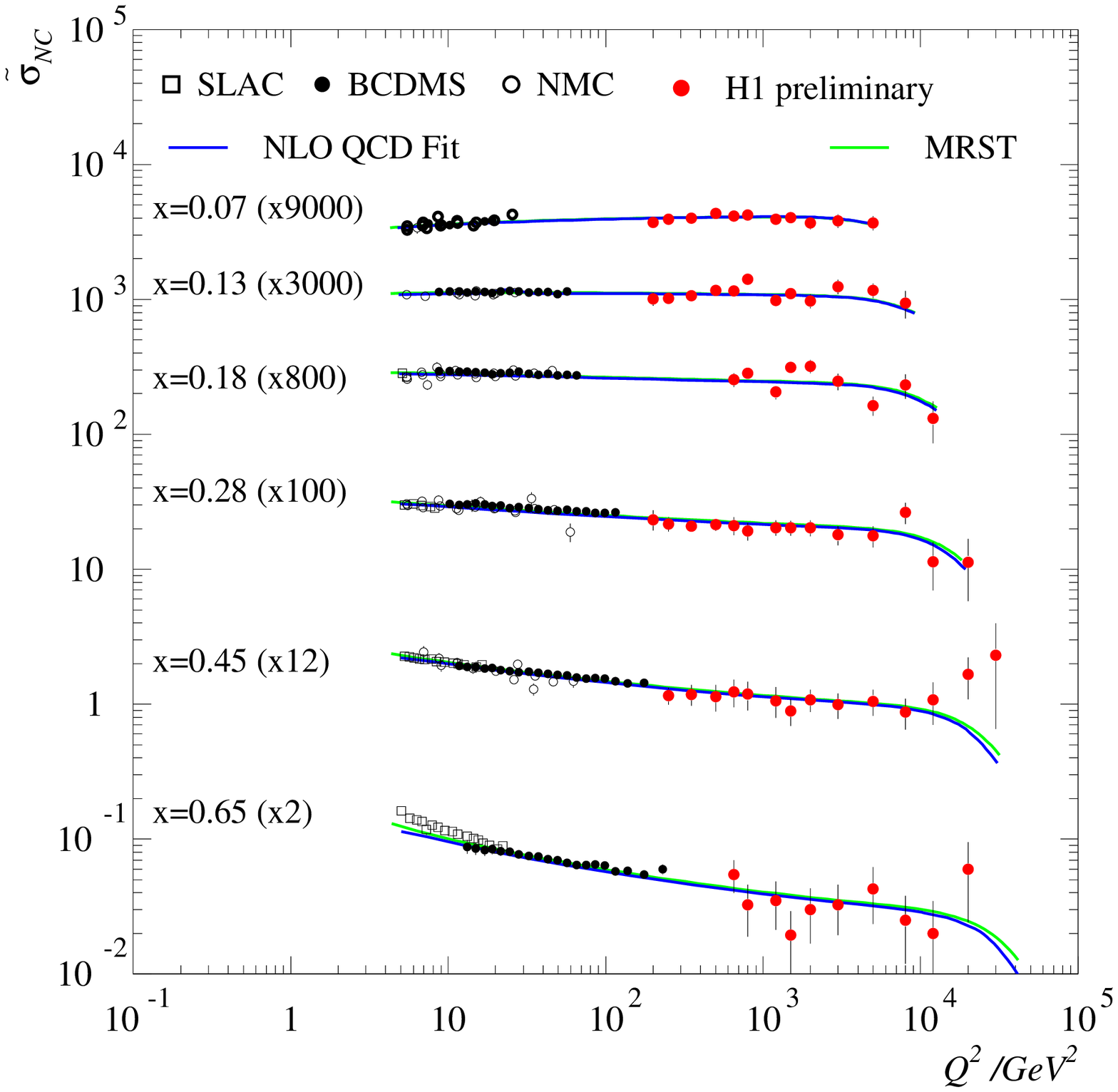}
  \end{center}
  \caption{
    \label{Fig:Fig3}
    The reduced cross section, as defined in equation 1,
    compared with the expectations of the Standard Model evaluated using
    the MRST \cite{MRST} PDFs and the result of the H1 NLO QCD
    fit. The error bars represent the total error of the measurements
    \protect{\cite{Vancouver}}.
  }
\end{figure}

A second motivation for significantly improving the NC data set at
high $x$ and high $Q^{2}$ is indicated in figure \ref{Fig:Fig3} where 
$\tilde{\sigma}^{\rm NC}_{e^{+}p}$ is plotted as a function of $Q^{2}$
for various  
fixed values of $x$ \cite{Vancouver}. For $x = 0.45$ the data points at the 
highest $Q^{2}$ lie above the SM expectation. In order to establish this 
effect as something other than a statistical fluctuation requires a 
significant increase in the size of the NC data set. 
\begin{figure}
  \begin{center}
    \includegraphics[width=.6875\textwidth]{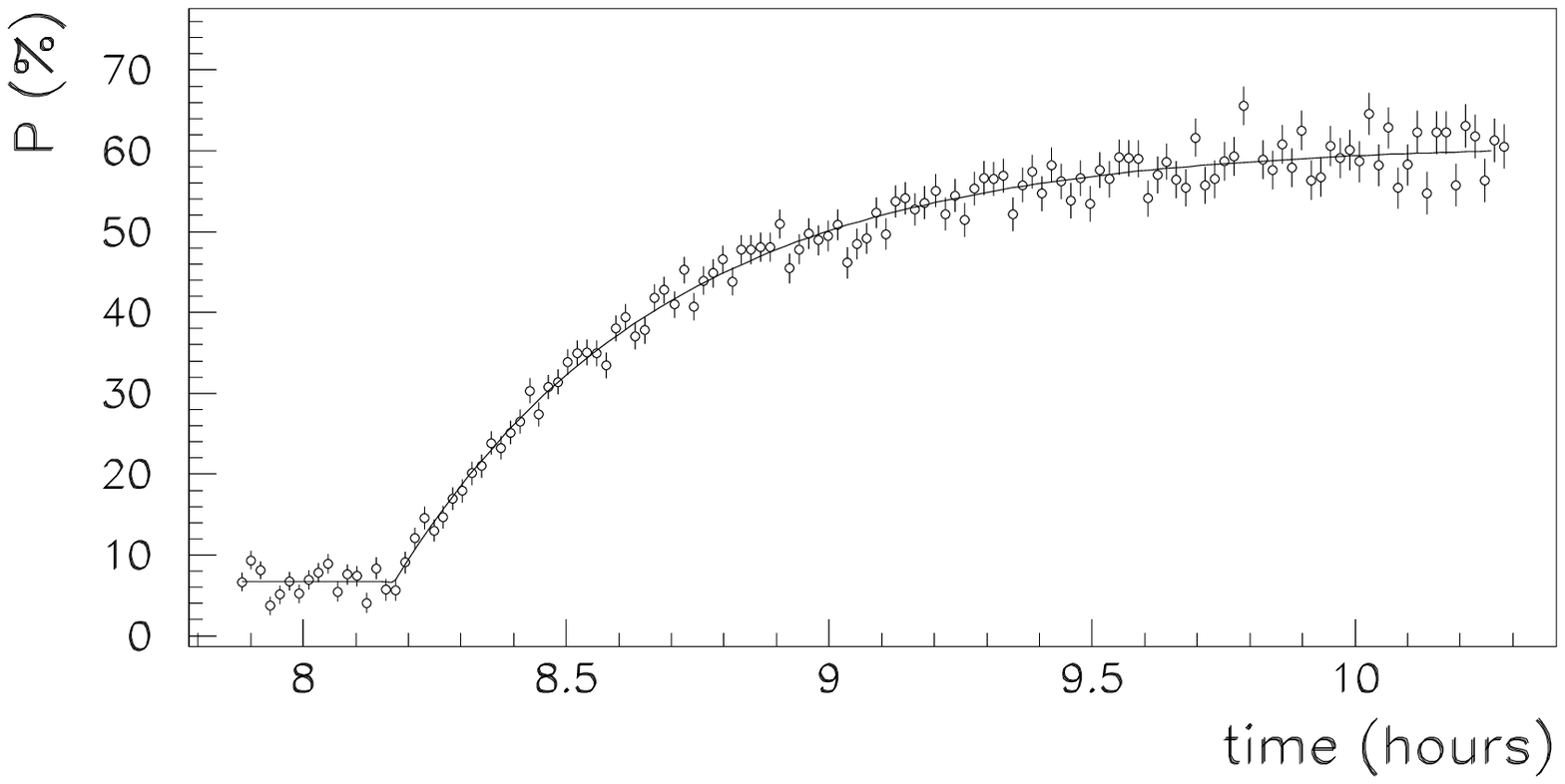}
  \end{center}
  \caption{
    \label{Fig:Fig4}
    A measurement of the build up of polarisation as a function
    of the time after the beam has been depolarised. The solid line shows
    the result of a fit to the data \protect{\cite{Andreev}}.
  }
\end{figure}

\subsection{Searches for Physics Beyond the Standard Model}

It is in the nature of searches for physics beyond the SM to push to
the edges of phase space. Hence, such analyses are always in need of a
large increase in data. Searches for physics beyond the SM at HERA are
reviewed elsewhere in these proceedings \cite{Kermiche}. The limits presented 
in reference \cite{Kermiche} indicate that HERA plays a crucial role in 
exploring the full panorama of new physics precisely because it is the worlds
only $ep$ collider. As such the HERA collider experiments have the
potential to make decisive contributions in the search for R-parity
violating SUSY, leptoquarks, contact interactions and much more
besides.

\section{The Upgrades to the HERA Machine and the Collider Detectors}

\subsection{The HERA Accelerator Upgrade}

The goals of the HERA upgrade programme are to provide an increase of
a factor of five in luminosity and to provide longitudinal lepton beam
polarisation for ZEUS and H1. Over a six year running period it is
anticipated that a total luminosity of 1000 pb$^{-1}$ will be delivered
\cite{Schneekloth}. 

The key parameters of the upgraded machine are summarised in table 1
\cite{Schneekloth}. The five fold increase in luminosity is to be achieved by
stronger focusing of the lepton and proton beams. In order to achieve
the strong focusing required superconducting magnets must be installed
close to the interaction region inside the H1 and ZEUS detectors. In
order to calculate the synchrotron radiation background a maximum
lepton beam energy, $E_{e}$, of 30 GeV was used. In operation $E_{e}$ will be
reduced somewhat to ensure that the RF system performs reliably at high 
current. HERA has been operating reliably with a proton beam energy of 920 
GeV since 1998 so that it is likely that the proton beam energy will remain 
920 GeV after the upgrade. 
\begin{table}
  \begin{center}
    \begin{tabular}{|l||c|c|}  \hline
      \multicolumn{3}{|c|}{HERA Parameters}  \\
      & 1997 & Upgrade \\ \hline
      p beam energy (GeV) & 820 & 820 \\
      e beam energy (GeV) & 27.5 & 30 \\
      Number of bunches & 180/189 & 180/189 \\
      Number of protons/bunch & $7.7 \times 10^{10}$ & $10 \times 10^{10}$ \\
      Number of electrons/bunch & $2.9 \times 10^{10}$ & $4.2 \times 10^{10}$ \\
      Proton current (mA) & 105 & 140 \\
      Electron current (mA) & 43 & 58 \\
      Hor. proton emittance (nm rad) & 5.5 & 5.7 \\
      Hor. electron emittance (nm rad) & 40 & 22 \\
      Proton beta function x/y (m) & 7/0.5 & 2.45/0.18 \\
      Electron beta function x/y (m) & 1/0.7 &0.63/0.26 \\
      beam size $\sigma_{x} \times \sigma_{y} (\mu m)$ & $200 \times 54$ & $118 \times 32$ \\
      Synchrotron Rad. at IP (kW) & 6.9 & 25 \\ \hline
      Specific luminosity (cm$^{-2}$ s$^{-1}$ mA$^{-2}$) & $7.6 \times 10^{29}$ & $1.4 \times 10^{30}$ \\
      Luminosity (cm$^{-2}$ s$^{-1}$) & {\bf $1.5 \times 10^{31}$} & {\bf $7 \times 10^{31}$} \\ \hline
    \end{tabular}
  \end{center}
  \caption{
    \label{Tab:Tab1}
    HERA parameters achieved in 1997 compared to the specifications of the luminosity upgrade
    programme \cite{Schneekloth}. 
  }
\end{table}

The second major goal of the HERA upgrade, the provision of
longitudinal lepton beam polarisation for the collider experiments,
will be achieved by the provision of spin rotators for ZEUS and
H1. Figure 4 shows the build up of polarisation during HERA running in
1998 \cite{Andreev}. The asymptotic value of $\sim$60\% is routinely
obtained in 
HERA. Spin rotators have been operating at the HERMES interaction
point for several years and are routinely providing 60\% longitudinal
polarisation. The present performance falls only slightly short of the
design goal of 70\% lepton beam polarisation at HERA after the upgrade. 

\subsection{The Collider Detector Upgrades}

The collider detector upgrades focus on providing optimal performance
for charm and beauty tagging and for the reconstruction of the
scattered lepton at high $x$ and high $Q^{2}$ \cite{MVD,PRC}. 
Charm tagging is strongly enhanced by the ability to identify displaced 
vertices. Hence ZEUS will install a silicon micro-vertex detector. The boost 
of the lepton-quark centre of mass system throws the decay products of
charmed particles preferentially into the forward
direction. Hence the ZEUS micro-vertex detector will be equipped with
a set of `wheels' to cover the forward region. H1 already operates a
micro-vertex detector. In order to enhance the charm tagging
efficiency a forward silicon tracker will be installed and a set of
`wheels' will be added to the backward silicon tracker in order to
measure the azimuthal coordinate so complementing the existing `wheels'
which measure the radial coordinate. To further enhance track
reconstruction in the forward direction both collaborations will
upgrade the forward tracking system \cite{Straw,H1}. H1 will install 
a system of planar drift chambers while ZEUS is building a set of planar straw
tube trackers. These detectors will significantly enhance track
reconstruction in the forward direction and so lead to improved charm
tagging and enhanced reconstruction of the scattered lepton
particularly at high $x$ and at high $Q^{2}$ . 

In view of the redesigned lepton and proton beams both experiments are 
planning upgrades to the luminosity measurement systems and H1 plans to 
upgrade the system of roman pots used to detect elastically scattered 
protons. Both collaborations plan trigger upgrades in order to increase 
flexibility and selectivity, particularly for events containing charmed 
particles.

\section{Physics at HERA after the Upgrade}

Following the HERA upgrade the proton will be probed using each of the
four possible combinations of lepton beam charge and polarisation. The
combination of high luminosity and polarisation will lead to a rich
and diverse programme of measurements which can only be sketched below
using a few examples. A detailed description of the physics
opportunities afforded by the upgrade is to be found in reference
\cite{Ingelman}. 

\subsection{Proton Structure}

The large data volume will allow $F_{2}^{\rm NC}$ to be extracted with an 
accuracy of $\sim$3\% over the kinematic range $2 \times 10^{-5}<x<0.7$ 
and $2 \times 10^{-5} <Q^{2} <5 \times 10^{4}\,{\rm GeV}^{2}$ \cite{Botje}. 
This estimate is based on an assumed luminosity of 1000 pb$^{-1}$ and 
reasonable assumptions for systematic uncertainties. Such a measurement will 
represent a significant improvement over current results. If QCD evolution 
codes which go beyond next to leading order become available and a careful
study of the dependence of the systematic errors on the kinematic
variables is made it will be possible to determine $\alpha_{\rm S}$ from the 
scaling violations of $F_{2}^{\rm NC}$ with a precision of $\leq 0.003$. The 
gluon distribution will also be extracted from such a fit. The result of a 
fit in which systematic errors are handled in an optimal way and assuming a 
luminosity of 1000 pb$^{-1}$ is shown in figure 5. The gluon distribution 
will be determined with a precision of $\sim$ 3\% for $x = 10^{-4}$ and 
$Q^{2} = 20\,{\rm GeV}^{2}$
\begin{figure}
  \begin{center}
    \includegraphics[width=.55\textwidth]{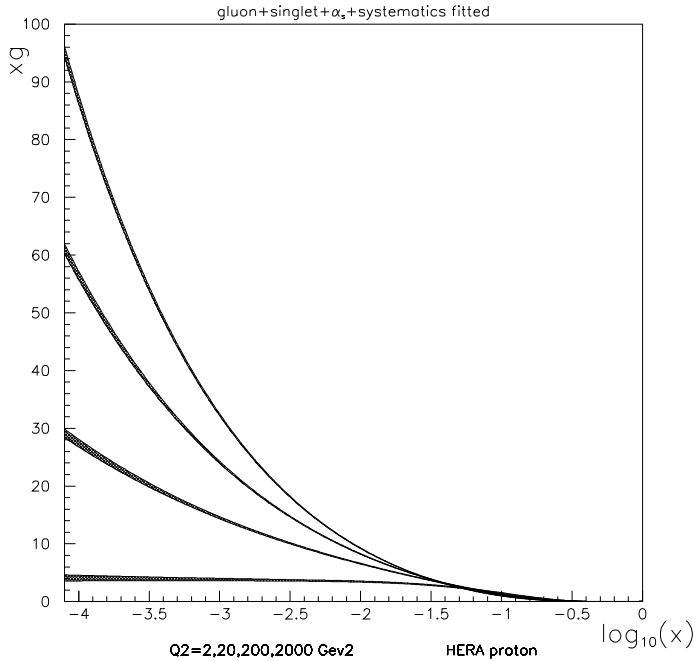}
  \end{center}
  \caption{
    \label{Fig:Fig5}
    Anticipated precision of a determination of the gluon density using
    a luminosity of 1000~pb$^{-1}$. Reasonable estimates of systematic 
    uncertainties have been included in the fit \protect{\cite{Botje}}.
  }
\end{figure}

The combination of high luminosity and high charm tagging efficiency 
transforms the measurement of the charm contribution to 
$F_{2}^{\rm NC}, F_{2}^{cc}$ \cite{Daum}. For example, figure 6 shows the 
precision on $F_{2}^{cc}$ expected from a luminosity of 500 pb$^{-1}$. The 
precision will be sufficient to allow a detailed study of the charm 
production cross section to be made. The lifetime tag provided by the silicon 
micro-vertex detector allows the tagging of $b$-quarks. Figure 7 shows the 
anticipated result of a measurement of the ratio of the beauty contribution to 
$F_{2}^{\rm NC}$, $F_{2}^{bb}$, to $F_{2}^{cc}$ assuming a luminosity of 
500 pb$^{-1}$ \cite{Daum}. The figure indicates that H1 and ZEUS will
be sensitive to the beauty content of the proton as well as the charm
content.  
\begin{figure}
  \begin{center}
    \includegraphics[width=.9375\textwidth]{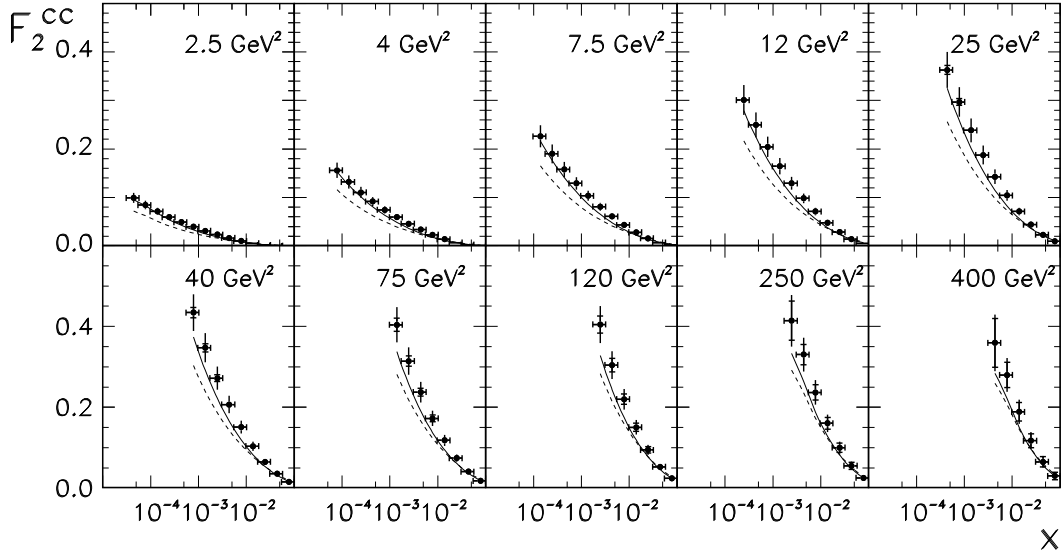}
  \end{center}
  \caption{
    \label{Fig:Fig6}
    Expected $F_{2}^{cc}$ for a luminosity of 500 pb$^{-1}$. The 
    inner (outer) error bars show the statistical (full) error of the anticipated
    measurement. The full (dashed) line gives the expectation from the NLO
    calculations based on GRV-HO \protect{\cite{GRVHO}} (MRSH
    \protect{\cite{MRSH}}) parton distributions taking a charm quark
    mass of 1.5 GeV \protect{\cite{Daum}}. 
  }
\end{figure}
\begin{figure}
  \begin{center}
    \includegraphics[width=.9375\textwidth]{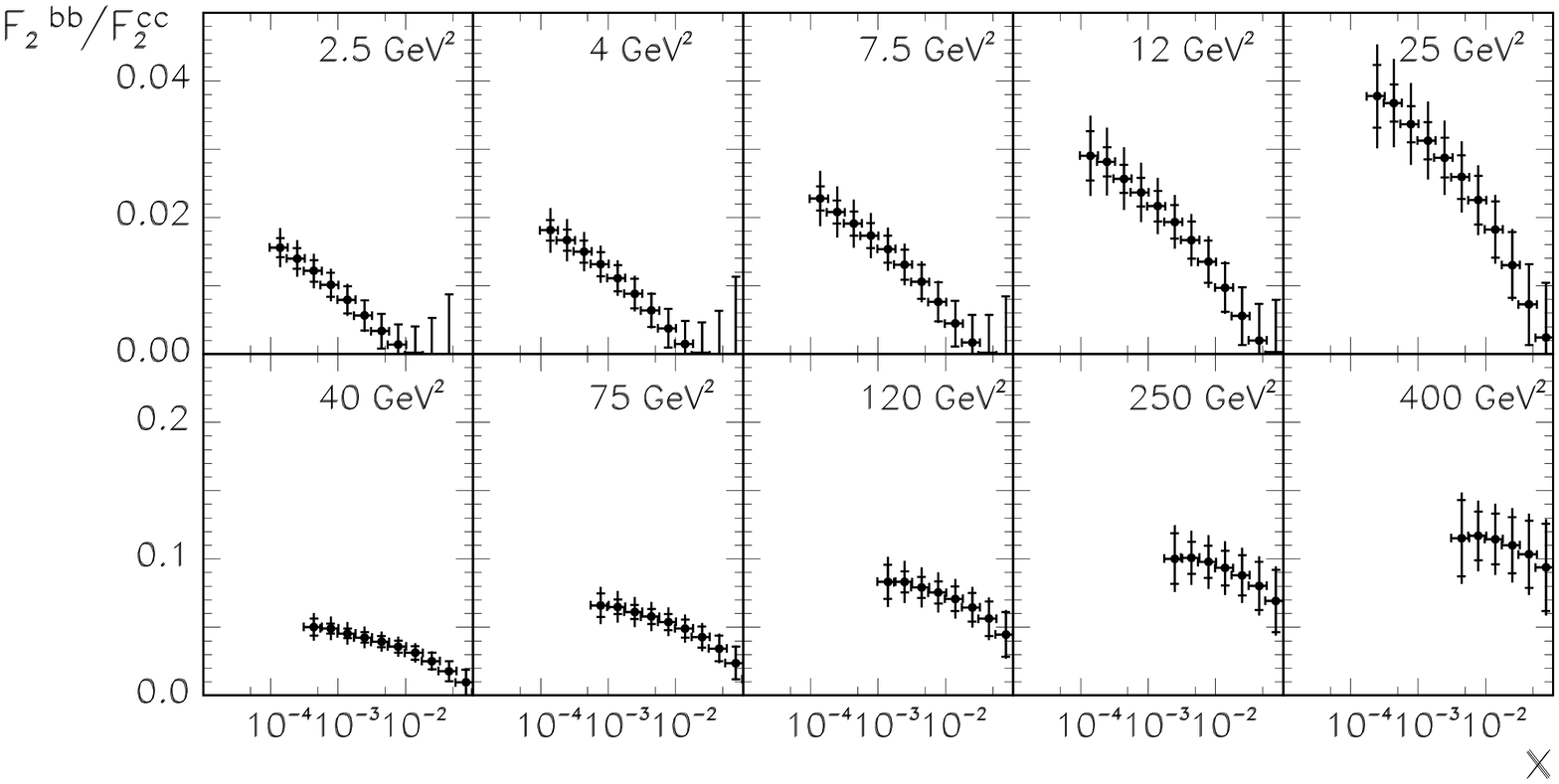}
  \end{center}
  \caption{
    \label{Fig:Fig7}
    Expected ratio of $F_{2}^{bb}$ to $F_{2}^{cc}$ for a 
    luminosity of 500 pb$^{-1}$. The inner (outer) error bars bars show the 
    statistical (full) error of the anticipated measurement
    \protect{\cite{Daum}}.
  }
\end{figure}

In the quark parton model CC DIS is sensitive to specific quark
flavours. The $e^{+}p$ CC DIS cross section is sensitive to the $d$- and 
$s$-quark parton densities and the $\bar{u}$- and $\bar{c}$-anti-quark 
densities, while the $e^{-}p$ CC DIS cross section is sensitive to the 
$u, c, \bar{d}$ and $\bar{s}$ parton density functions. With the large CC 
data sets expected following the upgrade it will be possible to use 
$e^{\pm}p$ CC data to determine the $u$- and $d$- quark densities. Further, by 
identifying charm in CC DIS it will be possible to determine the strange 
quark contribution to the proton structure function $F_{2}^{\rm NC}$ with an 
accuracy of between 15\% and 30\% \cite{Lamberti}.

In summary, following the upgrade the HERA collider experiments will
make a complete survey of the parton content of the proton. 

\subsection{Tests of the Electroweak Standard Model}

The high luminosity provided by the upgrade will allow access to low cross
section phenomena such as the production of real $W$-bosons. The SM
cross section for the process $ep \rightarrow eWX$ is $\sim$~1~pb 
\cite{WCross} which, combined with an acceptance of $\sim30\%$, gives a 
sizeable data sample for a luminosity of 1000 pb$^{-1}$. The production of 
the $W$-boson at HERA is sensitive to the non-abelian coupling $WW\gamma$ 
\cite{Noyes}. The sensitivity of HERA to non-SM couplings is comparable to 
the sensitivities obtained at LEP and at the Tevatron and complementary in 
that at HERA the $WW\gamma$ vertex is probed in the space-like regime. 

The full potential of electroweak tests at HERA will be realised
through measurements using polarised lepton beams \cite{Cashmore}. Two types of
electroweak test have been considered. The first involves the
interpretation of NC and CC cross section measurements in terms of
parameters of the SM such as the mass of the $W$ boson, $M_{W}$, and the mass
of the top quark, $m_{t}$. The consistency of the SM requires that the values
extracted must be in agreement with those obtained in measurements of
the same parameters in other experiments. The second form of SM test
involves the determination of parameters, such as the light quark NC
couplings, which are not free parameters in the SM. In this case a
deviation from the SM prediction would be a signal for new physics.

Within the SM NC and CC DIS cross sections may be written in terms of
$\alpha$, $M_{W}$ and $m_{t}$ together with the mass of the $Z$ boson, 
$M_{Z}$, and the mass of the Higgs boson, $M_{H}$. In order to test the 
consistency of the theory we may fix the values of $\alpha$ and $M_{Z}$ to 
those obtained at LEP or elsewhere and use measurements of the NC and CC DIS 
cross sections to place constraints in the $M_{W}$, $m_{t}$ plane for fixed 
values of $M_{H}$. The SM is consistent if the values of the parameters 
$M_{W}$ and $m_{t}$ obtained agree with the values determined in other 
experiments. The result of such an analysis is shown in figure 8 where one 
standard deviation contours in the $M_{W}$, $m_{t}$ plane have been derived 
from anticipated measurements of NC and CC DIS in $e^{-}p$ scattering with 
an electron polarisation of $-70\%$ \cite{Beyer}. Combining NC and CC data 
corresponding to a luminosity of 1000 pb$^{-1}$ with a top mass measurement 
from the Tevatron with a precision of $\sim\pm5$ GeV yields a measurement of 
$M_{W}$ with an error of $\sim60$ MeV. 
\begin{figure}
  \begin{center}
    \includegraphics[width=.56\textwidth]{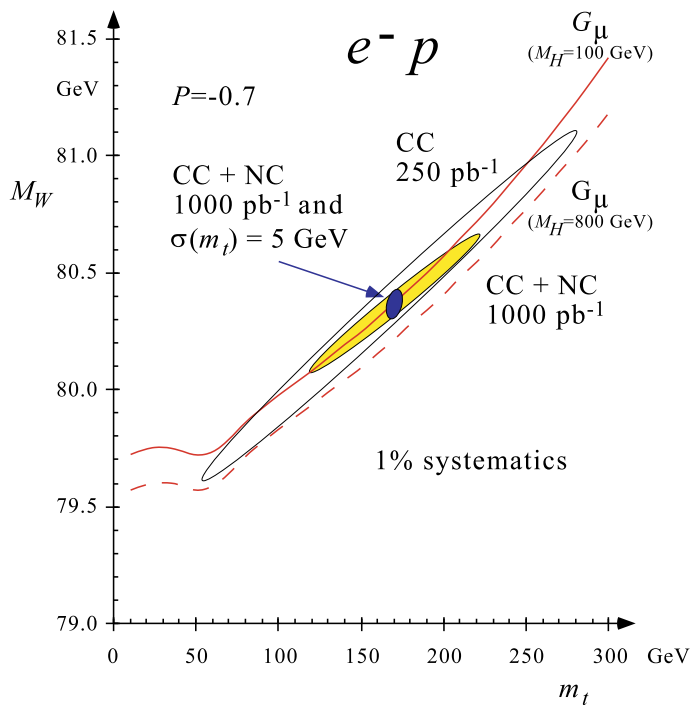}
  \end{center}
  \caption{
    \label{Fig:Fig8}
    One standard deviation contours in the $M_{W}$, $m_{t}$ plane 
    from polarised electron scattering ($P=-70\%$), utilising CC scattering at
    HERA alone with a luminosity of 250 pb$^{-1}$ (large ellipse), NC and CC
    scattering at HERA with 1000 pb$^{-1}$ (shaded ellipse), and the
    combination of the latter HERA measurements with a direct top mass
    measurement with a precision of 5 GeV (full ellipse). 
    The $M_{W}$ - $m_{t}$ relation following from the $G_{\mu}$ constraint 
    is shown for two values of $M_{H}$ (full and dashed curves)
    \protect{\cite{Beyer}}.
  }
\end{figure}
\begin{figure}
  \begin{center}
    \includegraphics[width=.75\textwidth]{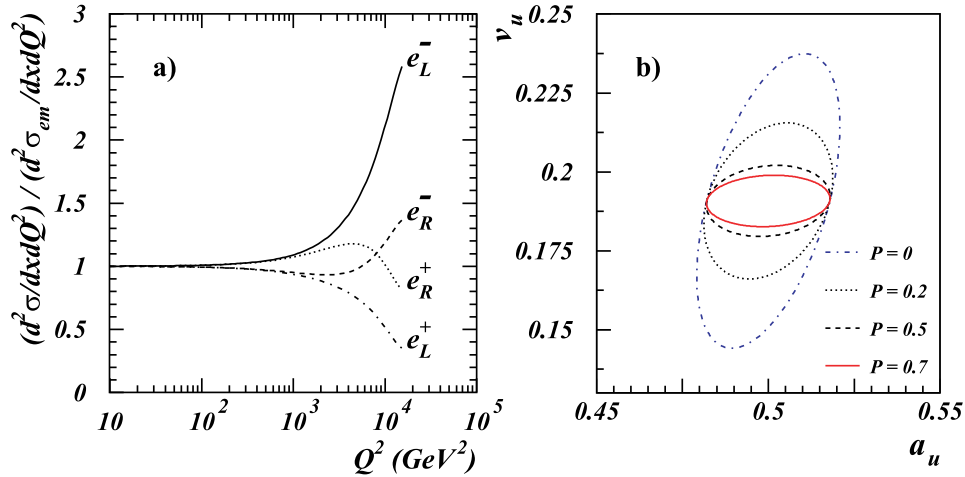}
  \end{center}
  \caption{
    \label{Fig:Fig9}
    (a) Ratio of the NC DIS cross section to the cross section obtained
    when only single photon exchange is included as a function of $Q^{2}$ 
    at $x=0.2$. (b) Sensitivity of the errors on the $u$-quark
    coupling to the $Z$ to the lepton beam polarisation $P$. One
    standard deviation contours are shown for fits in which the
    $u$-quark couplings are allowed to vary while the $d$-quark
    couplings are held fixed at their SM values
    \protect{\cite{Klanner}}. 
  }
\end{figure}

The sensitivity of NC DIS to lepton beam polarisation is shown in
figure 9(a). The figure shows the ratio  
\begin{equation}
R =\left( \frac{d^{2}\sigma^{\rm NC}} {dxdQ^{2}}\right)/
\left(\frac{d^{2}\sigma^{\rm em}}{dxdQ^{2}} \right) 
\end{equation}
where \( d^{2}\sigma^{\rm em}/dxdQ^{2} \) is the differential cross section 
obtained if only photon exchange is taken into account. The strong 
polarisation dependence of the NC cross section can be used to extract the NC 
couplings of the light quarks. In such an analysis the CC cross section may 
be used to reduce the sensitivity of the results to uncertainties in the
PDFs \cite{Klanner}. 
The precision of the results obtained depend strongly on the
degree of polarisation of the lepton beam as shown in figure 9(b). The
figure shows the anticipated error on the vector and axial-vector
couplings of the $u$-quark, $v_{u}$ and $a_{u}$ respectively, obtained in a 
fit in which $v_{u}$ and $a_{u}$ are allowed to vary while all other 
couplings are fixed at their SM values. With a luminosity of 250 pb$^{-1}$ 
per charge, polarisation combination and taking the vector and axial-vector
couplings of the $u$- and $d$-quarks as free parameters gives a precision
of 13\%, 6\%, 17\% and 17\% for $v_{u}$, $a_{u}$, $v_{d}$ and $a_{d}$ 
respectively. By comparing these results with the NC couplings of the $c$- and 
$b$-quarks obtained at LEP a stringent test of the universality of the NC 
couplings of the quarks will be made. 

\section{Summary}

Deep inelastic scattering has played, and continues to play, a central
role in the development of the understanding of the interactions among
the fundamental particles. The HERA upgrade programme provides
exciting opportunities which are both qualitatively and quantitatively
new. The measurements to be performed in the years following the HERA
upgrade will impinge directly on the description of the structure of
the proton, the nature of the strong interaction and the electroweak
sector of the Standard Model.


\newpage
\begin{thebibliography}{99}

\bibitem{Abt} H1 Collaboration, I. Abt et al., Nucl. Phys. {\bf B407} (1993) 515; \\
              H1 Collaboration, T. Ahmed et al., Nucl. Phys. {\bf B439} (1995) 471.

\bibitem{Derrick} ZEUS Collaboration, M. Derrick et al., Phys. Lett. 
{\bf B316} (1993) 412;\\
ZEUS Collaboration, M. Derrick et al., Z. Phys. {\bf C65} (1995) 379;\\
ZEUS Collaboration, M. Derrick et al., Z. Phys. {\bf C72} (1996) 399.

\bibitem{Cooper} A.M. Cooper-Sarkar, R. Devenish and A. de Roeck, Int'l
   J. Mod. Phys. {\bf A13} (1998) 3385. 

\bibitem{Abramovicz} H. Abramovicz, A. Caldwell, DESY-98-192 (December 1998).

\bibitem{Ingelman} Proc. of the Workshop ``Future Physics at HERA'', Vols. 1 
and 2, Eds. G. Ingelman, A. de Roeck, R. Klanner, DESY (1996). 

\bibitem{Ruckl} G. Ingelman and R. R$\ddot{\rm u}$ckl, Phys. Lett. {\bf B201} 
(1988) 369.

\bibitem{Aid} H1 Collaboration; S. Aid et al., Nucl. Phys. {\bf B470} (1996) 
4;\\
H1 Collaboration, C. Adloff et al., Nucl. Phys. {\bf B497} (1997) 3;\\
H1 Collaboration, Abstracts 534 and 535 contributed to the XXIX
International Conference on High Energy Physics, Vancouver, 23-29 July
1998. 

\bibitem{Breitweg} ZEUS Collaboration, J. Breitweg et al., Phys. Lett. 
{\bf B407} (1997) 432;\\ 
ZEUS Collaboration, M. Derrick et al., Z. Phys. {\bf C69} (1996) 607;\\
ZEUS Collaboration, M. Derrick et al., Z. Phys {\bf C72} (1996) 399;\\
ZEUS Collaboration, Abstract 770 contributed to the XXIX International
Conference on High Energy Physics, Vancouver, 23-29 July 1998. 

\bibitem{ZEUS} ZEUS Collaboration; J. Breitweg et al., Euro. Phys. J. {\bf C7} 
(1999) 609.

\bibitem{67} ZEUS Collaboration, J. Breitweg et al., Euro. Phys. J. {\bf C6} 
(1999) 67;\\
 ZEUS Collaboration, J.Breitweg et al., Phys. Lett. {\bf B407} (1997) 402.

\bibitem{Adloff} H1 Collaboration, C. Adloff et al., DESY-98-204, submitted to 
Nucl. Phys. B. 

\bibitem{DESY-99-056} ZEUS Collaboration; J. Breitweg et al.,
  DESY-99-056.

\bibitem{CTEQ4} H.L. Lai et al., Phys. Rev. {\bf D55} (1997), 1280.

\bibitem{Chekelian} V. Chekelian, `Neutral and Charged Currents at High 
$Q^{2}$ in $ep$ Interactions', invited paper at $34^{th}$ Rencontres de 
Moriond: Electroweak Interactions and Unified Theories, 13-20 March 1999,
 Les Arcs, France. 

\bibitem{MRST} A.D. Martin et al., Eur. Phys. J. {\bf C4} (1998) 463.

\bibitem{Vancouver} H1 Collaboration, Abstract 533 contributed to the XXIX 
International Conference on High Energy Physics, Vancouver, 23-29 
July 1998. 

\bibitem{Kermiche} S. Kermiche, `The Recent Beyond the Standard Model Results 
from the H1 and ZEUS (HERA) Experiments' these proceedings. 

\bibitem{Schneekloth} U. Schneekloth (editor), `The HERA Luminosity Upgrade', 
DESY-HERA-98-05, July 1998. \\ 
U.Scheekloth, `Recent HERA Results and
Future Prospects' DESY 98-060 and Moriond 1998.

\bibitem{Andreev} Polarisation 2000 Group, V. Andreev et al. DESY-PRC-98-07 
and DESY-PRC-99-05. 

\bibitem{MVD} ZEUS Collaboration, `A Micro Vertex Detector for ZEUS', 
DESY-PRC-97-01. 

\bibitem{PRC} H1 Collaboration, `A Forward Silicon Tracker for H1',
DESY-PRC-99-01. 

\bibitem{Straw} ZEUS Collaboration, `A Straw Tube Tracker for ZEUS', 
DESY-PRC-97-01. 

\bibitem{H1} H1 Collaboration, `Proposal for an Upgrade of the H1
Forward Track Detector for HERA 2000', DESY-PRC-98-06.

\bibitem{Botje} M. Botje, M. Klein, C. Pascaud, `Future Precision 
Measurements of $F_{2}(x,Q^{2}), \alpha_{S}(Q^{2})$ and $xg(x,Q^{2})$', Proc. 
Of the Workshop ``Future Physics at HERA", Vol. 1, Page 33, Eds. G. Ingelman, 
A. de Roeck, R. Klanner, DESY (1996). 

\bibitem{Daum} K. Daum et al., `The Heavy-Flavour Contribution to Proton 
Structure', Proc. Of the Workshop ``Future Physics at HERA", 
Vol. 1, Page 89, Eds. G. Ingelman, A. de Roeck, R. Klanner, DES (1996). 

\bibitem{GRVHO} M. Gl$\ddot{\rm u}$ck et al., Z. Phys. {\bf C53}
(1992) 127.

\bibitem{MRSH} P.N. Harriman et al., Phys. Rev {\bf D42} (1990) 798.

\bibitem{Lamberti} L. Lamberti, private communication.

\bibitem{WCross} ZEUS Collaboration, J. Breitweg et al., DESY-99-054,
  submitted to Phys. Lett. B.

\bibitem{Noyes} V.A. Noyes, `Limits on the $WW\gamma$ Couplings from Single $W$ 
Boson Production in $ep$ Collisions', Proc. Of the Workshop ``Future 
Physics at HERA", Vol. 1, Page 190, Eds. G. Ingelman, A. de Roeck, 
R. Klanner, DESY (1996). 

\bibitem{Cashmore} R. Cashmore et al., `Electroweak Physics at HERA: 
Introduction and Summary', Proc. Of the Workshop ``Future Physics at HERA", 
Vol. 1, Page 129, Eds. G. Ingelman, A. de Roeck, R. Klanner, DESY (1996). 

\bibitem{Beyer} R. Beyer et al., `Electroweak Precision Tests with Deep 
Inelastic Scattering at HERA', Proc. Of the Workshop ``Future Physics at 
HERA", Vol. 1, Page 140, Eds. G. Ingelman, A. de Roeck, 
R. Klanner, DESY (1996). 

\bibitem{Klanner} R.J. Cashmore et al., `Measurement of Weak Neutral Current 
Couplings of Quarks at HERA', Proc. Of the Workshop ``Future 
Physics at HERA", Vol. 1, Page 163, Eds. G. Ingelman, A. de Roeck, 
R. Klanner, DESY (1996). 

\end{thebibliography}
\end{document}